# Application of data engineering approaches to address challenges in microbiome data for optimal medical decision-making

Isha Thombre[1], Pavan Kumar Perepu[1], Shyam Kumar Sudhakar[1*]

[1]Krea University, 5655, Central Expressway, Sri City, Andhra Pradesh 517646

**\* Correspondence:** Shyam Kumar Sudhakar - Krea University, 5655, Central Expressway, Sri City, Andhra Pradesh 517646, India. Phone: +91 8015920041 Email: shyamkumar.sudhakar@krea.edu.in

**Running title**: Machine learning for personalized medicine using microbiome data

*Number of figures*: 3

*Number of tables*:  5

**Abstract**

The human gut microbiota is known to contribute to numerous physiological functions of the body and also implicated in a myriad of pathological conditions. Prolific research work in the past few decades have yielded valuable information regarding the relative taxonomic distribution of gut microbiota. Unfortunately, the microbiome data suffers from class imbalance and high dimensionality issues that must be addressed. In this study, we have implemented data engineering algorithms to address the above-mentioned issues inherent to microbiome data. Four standard machine learning classifiers (logistic regression (LR), support vector machines (SVM), random forests (RF), and extreme gradient boosting (XGB) decision trees) were implemented on a previously published dataset. The issue of class imbalance and high dimensionality of the data was addressed through synthetic minority oversampling technique (SMOTE) and principal component analysis (PCA). Our results indicate that ensemble classifiers (RF and XGB decision trees) exhibit superior classification accuracy in predicting the host phenotype. The application of PCA significantly reduced testing time while maintaining high classification accuracy. The highest classification accuracy was obtained at the levels of species for most classifiers. The prototype employed in the study addresses the issues inherent to microbiome datasets and could be highly beneficial for providing personalized medicine.

# 1. Introduction

The human body is home to trillions of microorganisms that contribute to both normal physiological functioning and disease (Fan and Pedersen 2021; Afzaal *et al.* 2022; Hou *et al.* 2022). The human microbiota, particularly those occupying the gut, influences the activity of multiple important organs of the body and is involved in several physiological functions (homeostasis, inflammation, and metabolism) (Fan and Pedersen 2021; Afzaal *et al.* 2022; Hou *et al.* 2022). In addition to that, the human microbiota is well documented to be implicated in numerous pathological conditions (Fan and Pedersen 2021; Afzaal *et al.* 2022; Hou *et al.* 2022). Dysbiosis of the gut microbiota is known to be implicated in inflammatory bowel disease (Matsuoka and Kanai 2015; Qiu *et al.* 2022), irritable bowel syndrome (Chey and Menees 2018), urinary tract infection (Worby *et al.* 2022), pulmonary (Shi *et al.* 2021), and cardiovascular abnormalities (Novakovic *et al.* 2020), and many other conditions (Guinane and Cotter 2013; Hayden *et al.* 2020). The gut microbiota also influences brain function and is known to be involved in myriad diseased brain states like Alzheimer's disease (Varesi *et al.* 2022) and other psychiatric (MacQueen *et al.* 2017; Andrioaie *et al.* 2022) and neurological (Cryan *et al.* 2020; Suganya and Koo 2020) conditions.

The past numerous years have seen immense research work on the physiology of the human gut microbiota (Huttenhower *et al.* 2012; McDonald *et al.*). Big microbiome projects through whole genome sequencing techniques have yielded valuable information on the form and function of the human microbiota. The resulting data (Holmes 2019; Kuntal and Mande 2019; Shetty and Lahti 2019) has motivated the machine learning community to come up with predictive models (Pasolli *et al.* 2016; Thomas *et al.* 2019; Vangay *et al.* 2019; Haque and Mande 2019; Marcos-Zambrano *et al.* 2021) to infer host phenotypes based on careful feature selection. The modeling efforts are assumed to play an important role in providing personalized medicine (Haque and Mande 2019) by creating a connection between the taxonomic makeup of the human microbiome and disease conditions and/or human health (Liñares-Blanco et al. 2022). In this regard, a number of machine learning studies have successfully predicted the host phenotypes (Thomas et al. 2019; Vangay et al. 2019; Hayden et al. 2020; Liñares-Blanco et al. 2022; Giuffrè et al. 2023) or disease states based on the taxonomic composition of the microbiome. Additionally, personalized medicine is also possible by developing novel therapeutics that can be utilized to target the microbiome in an individualized manner (Liñares-Blanco et al. 2022).

However, the classification of host phenotypes using the gut microbiome of human subjects has its own challenges that could hamper providing personalized medicine. The challenges could be two-fold arising both from the nature of the human microbiome data and from the nuances of using machine learning algorithms. Human microbiome data are often multidimensional (Thompson et al. 2017a; Armstrong et al. 2022) and sparse (Haque and Mande 2019; Pan 2021) along with significant variability between the subjects (Huttenhower et al. 2012; Zhang et al. 2019; Kodikara et al. 2022). In addition to that, microbiome data is often unbalanced (Anyaso-Samuel et al. 2021; Díez López et al. 2022) leading to the underrepresentation of the minority class in the classification of host phenotypes. To add to the problems of the clinical microbiologist working on personalized medicine for human subjects, machine learning models often come with a myriad of parameters that could be complex to deal with. As a result of the above-mentioned challenges, machine learning based classification based on microbiome data might lead to below-par performance accuracies (Bokulich et al. 2022; Díez López et al. 2022) and models that work on one dataset might not necessarily perform accurately when applied to similar other ones (LaPierre et al. 2019; Wang and Liu 2020; Díez López et al. 2022). Therefore, in order to effectively utilize human microbiome information for providing personalized medicine, one needs to address the above-mentioned issues arising out of the nature of the microbiome data.

In this study, we have addressed the issues of the unbalanced and multidimensional nature of microbiome datasets using data engineering approaches like synthetic minority oversampling technique (SMOTE) (Chawla et al. 2002) and principal component analysis (PCA) (Jolliffe 2011) etc. After data preprocessing, we have chosen four standard machine learning classifiers (Logistic Regression (LR), Support Vector Machines (SVM), Random Forests (RF), and Extreme Gradient Boosting decision trees (XGB)) and compared their accuracies in inferring host phenotypes. We believe that this is the first attempt on addressing the issue of high dimensionality and class imbalances on the microbiome data for the classification of host phenotypes at multiple levels of taxonomic hierarchy.

The dataset that was utilized in the study captures the gut microbiome signatures of cystic fibrosis infants with normal and abnormal growth profiles (https://microbiomedb.org/mbio/app). With this approach, we aim to quantify the performance of the above-mentioned classifiers in predicting the host phenotype (low length vs normal length) after applying minority oversampling (Chawla *et al.* 2002) and feature reduction algorithms (Jolliffe 2011) to the microbiome data. In addition to that, the classification task was performed at multiple levels of taxonomic hierarchy to compare their relative accuracies across the classifiers employed in the study. Our results reveal that there is a significant heterogeneity in the performance of various classifiers especially when oversampling algorithms were applied to the dataset. The workflow adopted in this paper could act as a prototype for machine learning based classification to infer host phenotypes and/or disease states using the microbiome data.

## 2. Materials and Methods

The aim of the study is to create a machine learning workflow to address issues inherent to large-scale human microbiome datasets (Thompson *et al.* 2017a; Pan 2021; Anyaso-Samuel *et al.* 2021; Armstrong *et al.* 2022; Díez López *et al.* 2022). For this purpose, we employed dimensionality reduction (Jolliffe 2011) and oversampling algorithms (Chawla *et al.* 2002) to a previously published human microbiome dataset (Hayden *et al.* 2020) and studied the performance of the following machine learning classifiers: SVM, LR, RF, and XGB decision trees. The classification of host phenotype (low length vs normal length) was performed at multiple levels of taxonomic hierarchy (species, genus, family, class, order, phylum). The methodological details of the study are listed below in detail.

### 2.1 Microbiome data

The microbiome data utilized in the study was obtained from the website of the microbiome repository ("(https://microbiomedb.org/)"). The dataset ("https://microbiomedb.org/mbio/app/record/dataset/DS_fc61dff608") belongs to a former study that links fecal dysbiosis with atypical linear growth of infants with cystic fibrosis (Hayden *et al.* 2020). Three important files associated with the dataset were utilized in this study. They are listed as follows: file 1 (subject information file, supplementary information from the paper), file 2 (BONUS.WGS.sample detals.tsv, and file 3 (BONUS.WGS.taxon abundance.csv). Sequencing was done using shotgun metagenomic sequencing of the fecal samples of the subjects. Taxonomic classification of the microbiomes present in the samples was obtained using phylogenetic analysis post-sequencing (Bhat *et al.* 2019; Hayden *et al.* 2020). The study was exempted from IRB review as it involved secondary analysis of public use datasets with no identifying information.

Subject ID and the host phenotype (whether the subjects exhibited low length or normal length) were present in file 1. File 2 was used to map the information between file 1 and file 3. File 2, in addition to containing the subject ID, the age of the participants, also has the sample ID. The sample ID is a unique number that identifies the sample collected from a particular subject at a specific age. The sample ID in file 2 was used to retrieve microbiome taxonomic frequency information from file 3 which hosts the taxonomic frequency distribution of all microbiomes sequenced in the study. The information from the three files was finally consolidated into a single final dataset. In the final dataset, the total number of subjects (infants with cystic fibrosis) is 201. Of these 201 subjects, 153 had normal length growth profiles whereas 48 exhibited low length. Since each subject is represented multiple times in the dataset as their samples were collected at multiple time points, the total number of entries in the dataset that belong to normal and low length are 836 and 275 respectively (Table 1). Their distribution according to the various ages at sample collection can be seen in Table 1. At the species level, the dataset includes 643 features that represent taxonomic classification of microbiomes found in the fecal samples of the subjects. This number for genus, family, order, class and phylum are 213, 84, 37, 21, and 9 respectively. Furthermore, the age of the subjects at sample collection was added as a feature to aid classification. Due to sample size limitations, samples with age of 2 months or low were not considered for analysis. In (Hayden *et al.* 2020), the detailed composition of microbiomes has been described in detail.

| Age (months) | n | Normal | Low |
|---|---|---|---|
| 3 | 153 | 117 | 36 |
| 4 | 151 | 109 | 42 |
| 5 | 163 | 126 | 37 |
| 6 | 167 | 123 | 44 |
| 8 | 164 | 125 | 39 |
| 10 | 161 | 123 | 38 |
| 12 | 152 | 113 | 39 |
| Total | 1111 | 836 | 275 |

**Table 1**: Distribution of samples in the dataset according to age at sample collection and growth profile (normal vs low length) of the subjects. 'n' represents the number of samples, 'n' is the same for all levels of taxonomic hierarchy

## 2.2 Issues in Microbiome data

As mentioned earlier, there are some issues with the microbiome dataset: (a) Imbalanced, as the number of participants for the two classes of normal and low-length growth profiles is not uniform. Specifically, in the dataset, 836 samples belong to the majority class (normal length), while the minority class (low length) has only 275 samples (as shown in Table 1). (b) High dimensional (around 600 features/dimensions at species level).

For the imbalance issue, we sought to use the SMOTE algorithm (Synthetic Minority Oversampling Technique) (Chawla *et al.* 2002) to generate synthetic samples based on the number of nearest neighbors. To handle the high dimensionality problem, we applied principal component analysis (Jolliffe 2011) on the dataset to reduce the number of features. These data engineering/preprocessing approaches have been applied on the dataset and then fed to machine learning models.

## 2.3 Machine learning approaches

The machine learning models that were explored in this study are SVM, LR, RF, and XGB decision trees. Further, we compared the performance of the above classifiers before and after applying SMOTE and PCA. In the above data preprocessing as well as classification models, there are some hyperparameters to be considered and tuned to obtain better performance. Particularly, we have focused on the crucial hyperparameters that can achieve an improvement in accuracy. For example, SMOTE generates synthetic sample points based on the number of nearest neighbors, which is an important hyperparameter. Similarly, we have considered the regularization parameter (C) for SVM and *number of decision trees* for RF and XGB. The range of values that we have considered for the hyperparameters will be discussed in the next section.

# 3. Results

In Fig. 1, we have shown the workflow used in our study. As seen in this figure, we started with the unbalanced dataset at the 'species' level of taxonomic hierarchy and conducted different experiments using the earlier mentioned four classifiers. Then we generated a balanced dataset using the SMOTE algorithm. For the hyperparameter (*number of nearest neighbors*) in this algorithm, we have chosen the range of values (1 to 7 in steps of 2). As discussed in the previous section, we have also performed hyperparameter tuning for the classifiers. Hyperparameter tuning was performed on the unbalanced dataset and the model with the best accuracy metrics was utilized for further analysis with SMOTE and PCA. For SVM, we have tuned the regularization hyperparameter (C, by varying from 1 to 7, in steps of 2). Similarly, for RF and XGB, the hyperparameter, *number of decision trees*, was considered in the range of 5 to 200, in steps of 10, for tuning. For SVM, the model with C = 1 yielded the highest accuracy. For RF and XGB decision trees, the models with *number of decision trees* set to 15 and 145 respectively gave the highest accuracy. These models were retained for further analysis with SMOTE and PCA.

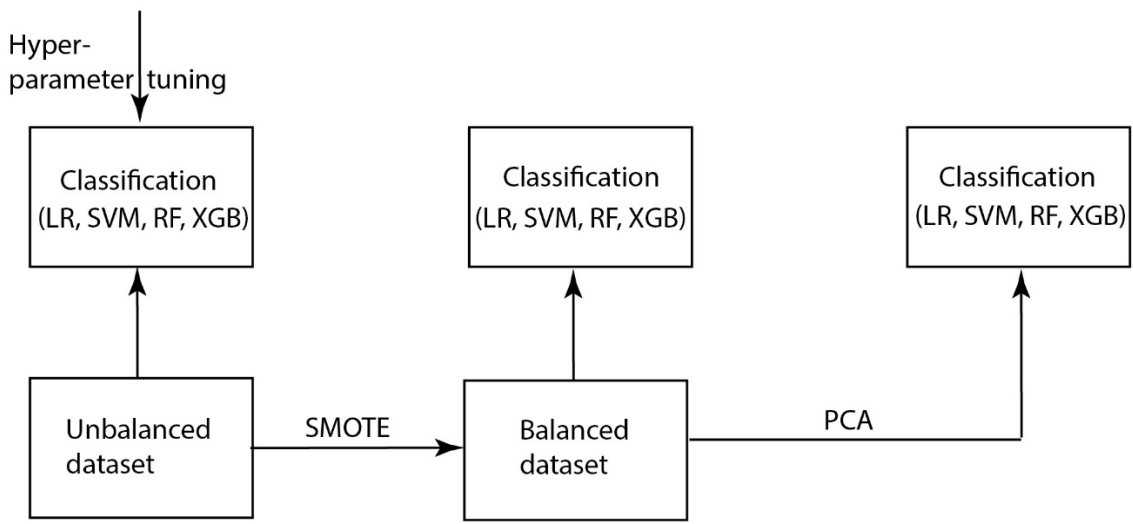

**Figure 1:** Workflow of algorithms implemented in the study: All four classifiers were first implemented on the unbalanced dataset. Hyperparameter tuning was performed on this dataset to select the model with the highest accuracy. The chosen model with the best accuracy was trained and tested with the balanced dataset generated by the SMOTE algorithm after removing the class imbalance. Further, PCA was implemented on the balanced dataset to further understand the tradeoff between performance and testing time of the classifiers.

Further, in our experimentation, we have also applied PCA on the balanced dataset and then fed the processed dataset (with a smaller number of features) to the classification models (four classifiers). If PCA is applied, unimportant features (with low variance) are eliminated (dimensionality reduction) and so testing time (classification/ prediction on test samples) is also reduced. After applying PCA, classification is based on only important features (but not on all of them) and so performance may be degraded. However, this experiment is helpful to figure out the tradeoff between the performance and testing time of the classifiers. PCA was applied only at 'species' level of taxonomy because the number of features was truly high dimensional (around 600 features/dimensions). This was not repeated for other levels of taxonomic hierarchy (genus, family, class, order, phylum) because the number of features/dimensions were far less compared to that of species.

For all the experiments, 80% of the data was utilized for training the classifiers and 20% for the purpose of testing and making predictions. For implementation of the above approaches, we used the scikit-learn package in Python programming language ("(https://scikit-learn.org/stable/)").

In Table 2, we have reported the performance of the classifiers with different workflows in terms of standard measures like accuracy, precision, and recall. These measures can be computed from a confusion matrix based on true (actual) and predicted values. The confusion matrix for the two-class classification problem (normal length vs low length in this study) is a 2 x 2 table that contains the following four values: true positives (TP), true negatives (TP), false positives (FP), and false negatives (FN). Performance measures like accuracy, precision, and recall can be computed from the above four values as shown below.

$$Accuracy = \frac{TN + TP}{TN + TP + FN + FP}$$

(1)

$$Precision = \frac{TP}{TP + FP}$$

(2)

$$Recall = \frac{TP}{TP + FN}$$

(3)

| Models | Accuracy | | | | | Precision | | | | | Recall | | | | |
|---|---|---|---|---|---|---|---|---|---|---|---|---|---|---|---|
| | UB | SM1 | SM3 | SM5 | SM7 | UB | SM1 | SM3 | SM5 | SM7 | UB | SM1 | SM3 | SM5 | SM7 |
| LR | 0.8 | 0.74 | 0.77 | 0.74 | 0.75 | 0.5 | 0.71 | 0.74 | 0.73 | 0.73 | 0.32 | 0.81 | 0.81 | 0.77 | 0.81 |
| SVM | 0.82 | 0.72 | 0.71 | 0.71 | 0.72 | 1 | 0.69 | 0.69 | 0.7 | 0.71 | 0.068 | 0.8 | 0.78 | 0.74 | 0.75 |
| RF | 0.83 | 0.86 | 0.85 | 0.85 | 0.87 | 0.82 | 0.88 | 0.86 | 0.85 | 0.89 | 0.2 | 0.84 | 0.84 | 0.84 | 0.84 |
| XGB | 0.82 | 0.89 | 0.9 | 0.89 | 0.9 | 0.6 | 0.9 | 0.92 | 0.9 | 0.9 | 0.34 | 0.87 | 0.87 | 0.87 | 0.9 |

**Table 2**: Accuracy, precision and recall of different machine learning classifiers implemented in unbalanced and balanced datasets. SM1, SM3, SM5 and SM7 refers to SMOTE algorithm with the hyperparameter, *number of neighbours*, set to 1,3,5 and 7 respectively. UB - unbalanced dataset

From Table 2, we can observe the following:

1. Using the unbalanced dataset, the accuracy was almost the same for all classifiers (approximately, 0.8). However, since the dataset is heavily unbalanced (836 samples from the normal length class and 275 samples from the low length class), all classifiers score poorly on the recall metric.

2. After applying the SMOTE algorithm with different values of hyperparameter (number of neighbors), varying from 1 to 7 in steps of 2, there is some marginal improvement in all the performance measure values for both RF and XGB decision trees classifiers.

3. However, after applying SMOTE, though there is an improvement in recall values, precision

values are reduced for SVM, and accuracies are marginally reduced for both LR and SVM. This may be due to the fact that both LR and SVM are linear classifiers based on a threshold/decision plane. SMOTE may generate synthetic samples in the proximity of the decision plane but on the wrong side which ultimately leads to their misclassification and so accuracy is reduced.

Since the microbiome dataset is sparse (Thompson *et al.* 2017a; Pan 2021; Armstrong *et al.* 2022) with a large number of features (high dimensional), we applied PCA and analyzed the tradeoff between performance and testing time. This analysis was performed on the balanced version of the dataset. In the PCA algorithm, we have retained only the features that cover 99% variance (Jolliffe 2011) and eliminated the remaining unimportant features with very low variances.

In Table 3, we have shown the accuracies after applying PCA along with the testing times (without and with PCA) for all the classifiers. Accuracies without PCA were shown in Table 2. For better illustration, we have shown accuracies without and with PCA in the form of plots in Fig. 2. It can be observed from Table 3 and Fig. 2, there is almost negligible degradation in performance. Interestingly, for SVM alone, there is an improvement in accuracy after applying PCA. It seems the elimination of features with low variance moved the synthetic samples to the correct side, thereby improving the accuracy. Of course, there is a significant reduction (by more than 50% and 80%), in testing time after PCA is applied (Fig. 3), for all the classifiers. Therefore, PCA can potentially be employed to handle the high dimensionality nature of microbiome datasets to bring down the classification time with no compromise on accuracy.

| Models | SM1 | | | SM3 | | | SM5 | | | SM7 | | |
|---|---|---|---|---|---|---|---|---|---|---|---|---|
| | $A$ | $T$ | $T_{PCA}$ | $A$ | $T$ | $T_{PCA}$ | $A$ | $T$ | $T_{PCA}$ | $A$ | $T$ | $T_{PCA}$ |
| LR | 0.76 | 12.3 | 1.57 | 0.76 | 13.68 | 1.35 | 0.76 | 17.75 | 1.9 | 0.77 | 12.78 | 1.5 |
| SVM | 0.85 | 190 | 187 | 0.82 | 196 | 198 | 0.79 | 244 | 209 | 0.83 | 193 | 194 |
| RF | 0.83 | 10.85 | 5.07 | 0.82 | 11.3 | 5.45 | 0.82 | 16.5 | 5.1 | 0.85 | 11.07 | 5.4 |
| XGB | 0.88 | 31.8 | 7.23 | 0.88 | 39.9 | 4.88 | 0.87 | 37.9 | 7.47 | 0.89 | 28.4 | 6.74 |

**Table 3**: Accuracy and testing time of different machine learning classifiers without and with PCA. T – testing time (without PCA), $T_{PCA}$ – testing time with PCA implemented on the dataset. SM1, SM3, SM5 and SM7 refers to SMOTE algorithm with the hyperparameter, *number of neighbours*, set to 1,3,5 and 7 respectively.

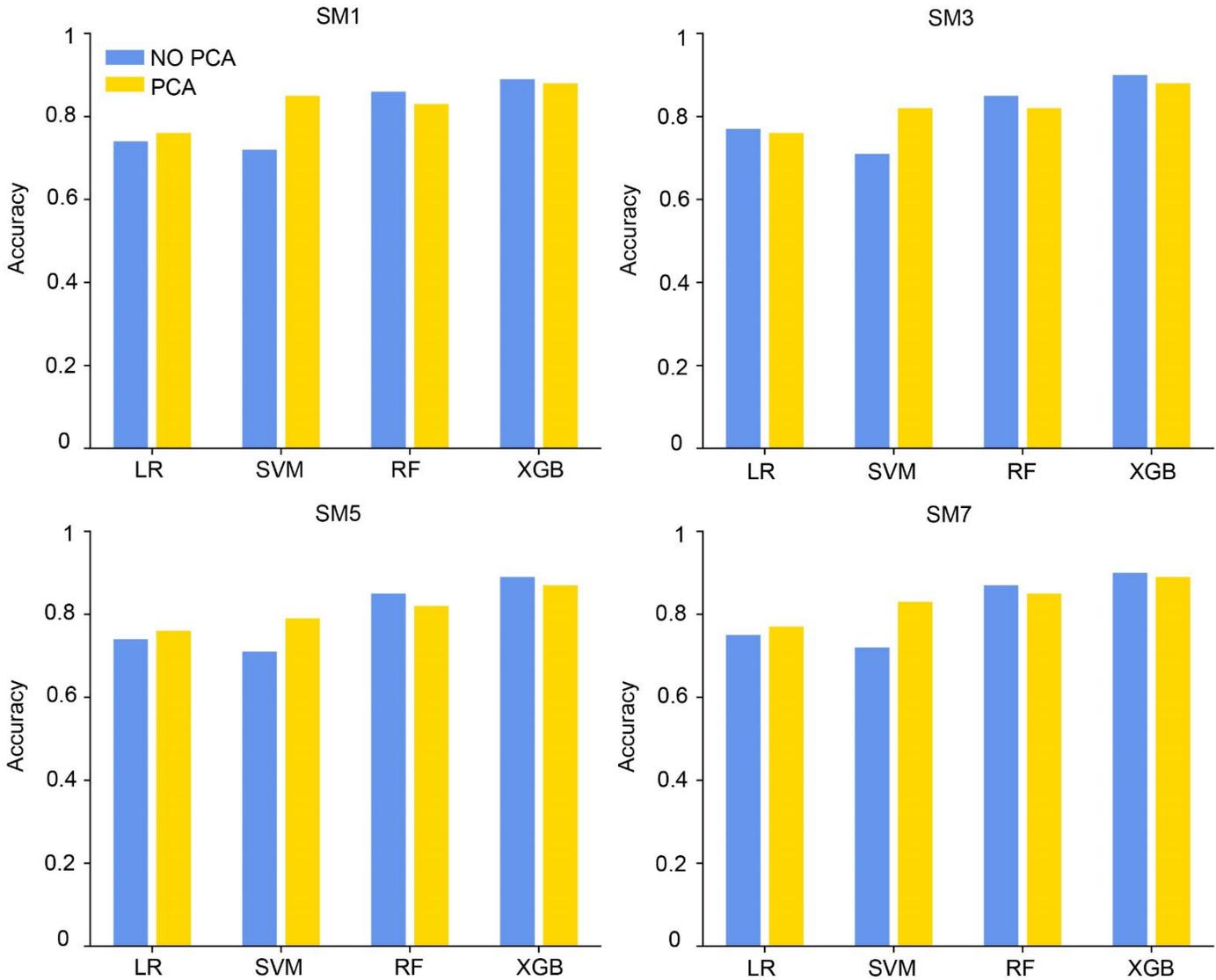

**Figure 2:** Accuracy of the classifiers with and without PCA implemented on the balanced dataset. SM1, SM3, SM5 and SM7 plots correspond to the SMOTE algorithm with 1, 3, 5, and 7 neighbors, respectively.

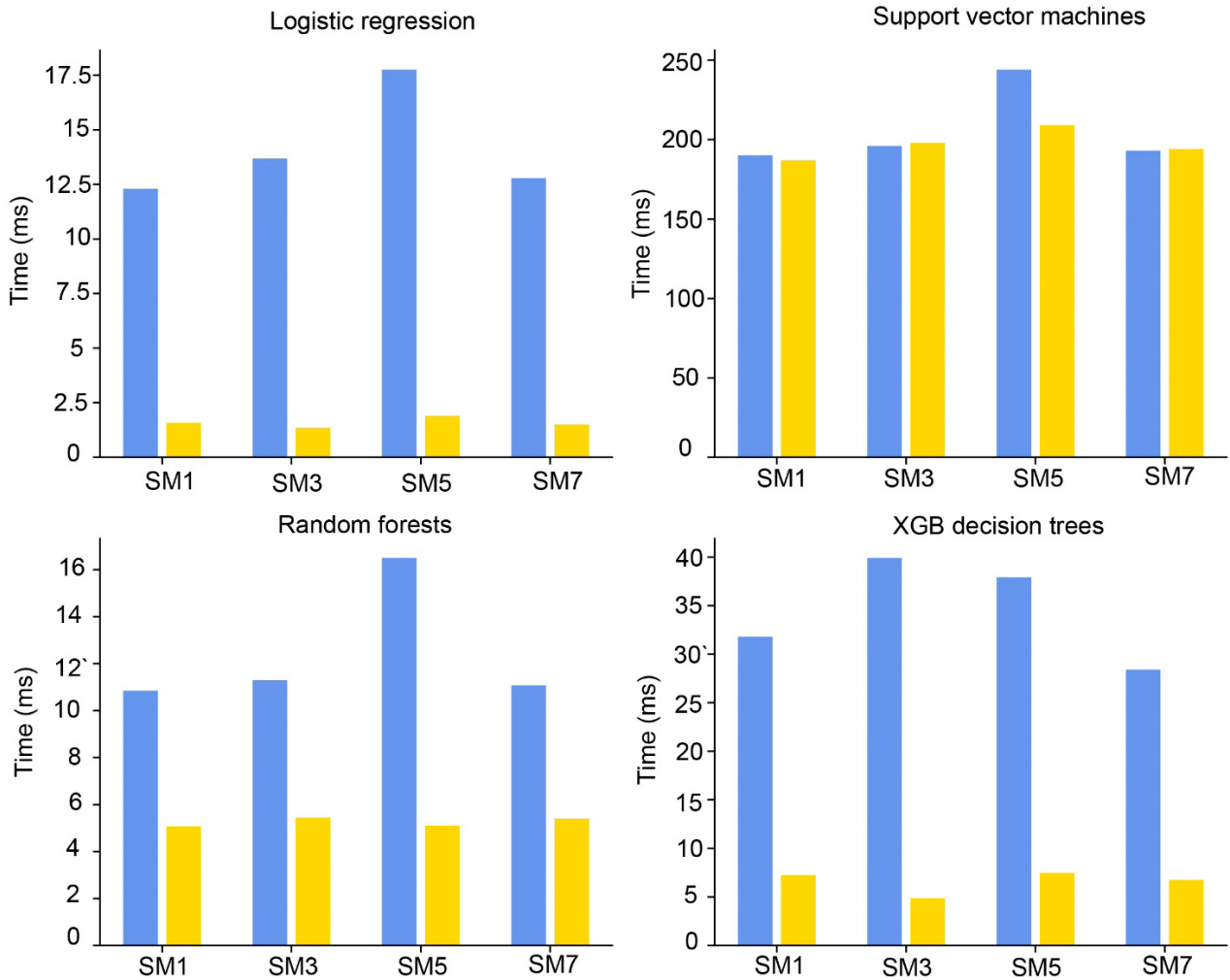

**Figure 3:** Testing time of the classifiers without and with PCA implemented on the balanced dataset. SM1, SM3, SM5 and SM7 plots correspond to the SMOTE algorithm with 1, 3, 5, and 7 neighbors, respectively.

We repeated the classification experiments for multiple levels of taxonomic hierarchy to understand the heterogeneity in the performance of the four classifiers at each level and to determine the taxon at which the highest classification performance is observed. As the number of features is not high, we have not applied PCA at other levels. Tables 4 and 5 show the classifier accuracies at multiple levels of taxonomic hierarchy for the unbalanced and balanced datasets respectively. From Tables 4 and 5, we can observe the following:

1. Similar to the classification at the level of the taxon 'species', the accuracies of all four classifiers were almost the same (around 0.8, Table 4) at the other levels of hierarchy (genus, family, order, class, and phylum) on the unbalanced dataset.
2. Similar to the taxon 'species', accuracies of SVM and LR dropped significantly on the balanced dataset generated using SMOTE (considering 5 neighbors, shown in Table 5).
3. From Table 5, we can see that across multiple levels of taxonomic hierarchy, the highest classification performance (accuracy) on the balanced dataset was obtained at the level of 'species' for the three classifiers, LR, SVM, and XGB. For only one classifier, RF, the highest performance on the balanced dataset was obtained at the level of 'genus'.
4. Among the four different classifiers, the highest classification accuracy was obtained for XGB decision trees at the levels of 'species' and 'genus' on the balanced dataset (Table 5).

| Models | Species | Genus | Family | Order | Class | Phylum |
|---|---|---|---|---|---|---|
| LR | 0.8 | 0.82 | 0.81 | 0.81 | 0.81 | 0.81 |
| SVM | 0.82 | 0.8 | 0.8 | 0.8 | 0.8 | 0.8 |
| RF | 0.83 | 0.81 | 0.81 | 0.8 | 0.78 | 0.78 |
| XGB | 0.82 | 0.8 | 0.78 | 0.78 | 0.79 | 0.79 |

**Table 4**: Accuracy of the different machine learning classifiers at multiple levels of taxonomical hierarchy on the unbalanced dataset

| Models | Species | Genus | Family | Order | Class | phylum |
|---|---|---|---|---|---|---|
| LR | 0.74 | 0.66 | 0.61 | 0.61 | 0.6 | 0.6 |
| SVM | 0.71 | 0.66 | 0.61 | 0.59 | 0.59 | 0.6 |
| RF | 0.85 | 0.89 | 0.86 | 0.83 | 0.8 | 0.82 |
| XGB | 0.89 | 0.85 | 0.81 | 0.73 | 0.77 | 0.75 |

**Table 5**: Accuracy of the different machine learning classifiers at multiple levels of taxonomical hierarchy on the balanced dataset created by SMOTE algorithm (with the hyperparameter, *number of neighbours*, set to the value of 5).

From the above results, we can conclude that data engineering approaches like SMOTE, PCA etc. can be used to address inherent issues in the microbiome datasets and achieve better classification performance.

# 4. Discussion

In this study, we implemented data engineering approaches to address some of the issues inherent to large-scale microbiome datasets when performing medical decision-making. Our results indicate the presence of significant heterogeneity in the performance of various machine learning classifiers explored in the study particularly when SMOTE was applied on the dataset to remove the class imbalance. Further, one can use dimensionality reduction to bring down the testing time without compromising the performance of the classifiers.

Microbiome datasets are often characterized by a class imbalance in that one classification class is overrepresented while the other is underrepresented (Anyaso-Samuel et al. 2021; Díez López et al. 2022). While it may not be feasible to address the issue of class imbalance by collecting data from more subjects belonging to the minority class, it's more realistic to solve this issue through oversampling techniques. The underrepresentation of the minority class might lead to the inferior performance of the classifiers, false classifications belonging to the majority class, and the inability of the classifiers to replicate the performance across multiple similar other datasets (Wang and Liu 2020; Bokulich et al. 2022). In a study involving the prediction of different phenotypes relating to the cervicovaginal environment, the authors report that the presence of class imbalance leads to compromise the accuracy of the classifiers (Bokulich et al. 2022). Wang et al (Wang and Liu 2020) in a systematic comparison of the performance of ensemble vs traditional classifiers in predicting the phenotypes or disease states across multiple human microbiome datasets, reported that the presence of class imbalance could lead to variation in the performance of the classifiers across datasets even for the same phenotype. Finally, in a review (LaPierre et al. 2019) that encompasses machine learning based prediction of disease states using microbiome sequencing data, the authors report that some classifiers might achieve spuriously high-performance metrics by predicting all test samples to the majority class.

In our study, we created a balanced dataset by implementing SMOTE, a data augmentation technique that eliminates class imbalances in the dataset by the generation of synthetic samples (Chawla et al. 2002). Upon application of SMOTE, we found that performance of some classifiers improved (ensemble classifiers) whereas SVM and LR exhibited a deterioration in performance. We decided to address class imbalance in our dataset by SMOTE algorithm as it's widely used for machine learning based classification both for medical and microbiome datasets (Chawla et al. 2002; Blagus and Lusa 2013). Other data augmentation approaches to rectify class imbalance include adaptive synthetic sampling (ADASYN) (Institute of Electrical and Electronics Engineers) and tree-based associative data augmentation (TADA) (Sayyari et al. 2019). While ADASYN is an extension of SMOTE which focuses on generating samples more along the boundary between the majority and minority class, TADA is a novel data augmentation algorithm that overcomes class imbalance by generating samples based on the phylogenetic relationship between the microbiomes. Future extension of the current study could involve the implementation of ADASYN and TADA to address the problem of class imbalance in microbiome datasets.

Another problem that comes with the microbiome datasets is their high dimensionality and sparseness. Both amplicon and shotgun metagenomic sequencing typically produce multifold microbiota counts or units associated with each sample in the study. This is due to the sheer abundance of microbiota that is typically present in the human body. Often this could result in the features outnumbering the sample counts by multifold (Thompson et al. 2017b). The microbiome samples are also considered to be sparse because numerous microorganisms present in one sample might be completely absent in others. The high dimensionality and sparseness pose real challenges as they could result in erroneous results arising out of complex statistical analysis. Thankfully, dimensionality reduction techniques could come to the rescue and offer meaningful analysis of complex microbiome datasets. In this study, we have incorporated the PCA algorithm, one of the most commonly utilized methods for dimensionality reduction for complex biological datasets. Future extensions of the study could incorporate other dimensionality reduction techniques such as uniform manifold approximation and projection (UMAP) (Armstrong *et al.* 2021), principal coordinate analysis (PCoA), and those which consider phylogenetic relationships between the microbiome species.

Our results indicate that ensemble classifiers (RF and XGB decision trees) display superior classification performance over the other two classifiers explored in the study (SVM and LR). Our results are in line with what is observed in the literature(Wang and Liu 2020) with respect to machine learning based classification using human microbiome data. In an extensive analysis, Wang et al(Wang and Liu 2020) compared the performance of various classifiers across numerous datasets and host phenotypes. Overall, ensemble classifiers (XGB decision trees and RF) exhibit better performance compared to SVM in the study.

## 5. Conclusion

Our study aims to address some of the important issues (class imbalance and high dimensionality) with respect to the use of microbiome datasets for medical decision-making using data engineering methods. Four important machine learning based classifiers were explored in this study: SVM, LR, XGB decision trees, and RF. We removed class imbalance using SMOTE and reduced the dimensions of the feature space using principal component analysis. Our results reveal that ensemble classifiers XGB decision trees and RF outperform the other classifiers employed in the study. Dimensionality reduction significantly reduces the computation time without compromising the performance of the classifier. Our study could act as a prototype for the classification of disease states or host phenotypes using the microbiome data in the field of personalized medicine.

**Author Contributions Statement**

**Isha Thombre**: Conceptualization, Methodology, Investigation, Writing - Review and Editing. **Pavan Kumar Perepu**: Conceptualization, Methodology, Software, Validation, Formal Analysis, Investigation, Resources, Data Curation, Writing - Original Draft, Writing - Review and Editing, Visualization, Supervision, Project Management, Funding Acquisition. **Shyam Kumar Sudhakar**: Conceptualization, Methodology, Software, Validation, Formal Analysis, Investigation, Resources, Data Curation, Writing - Original Draft, Writing - Review and Editing, Visualization, Supervision, Project Management, Funding Acquisition

**Author Disclosure Statement**

The authors declare that the research was conducted in the absence of any commercial or financial relationships that could be construed as a potential conflict of interest.

**Funding**

This work was supported by faculty grant from Krea University.